\documentclass[%
 reprint,
superscriptaddress,
 amsmath,amssymb,
 aps,
]{revtex4-1}
\usepackage{graphicx}% Include figure files
\usepackage{dcolumn}% Align table columns on decimal point
\usepackage{bm}% bold math
\usepackage[colorlinks=true, citecolor=blue, linkcolor=blue]{hyperref}
\usepackage{siunitx} % for typesetting units correctly 
\usepackage{xargs}
\usepackage{url}
\begin{document}

\title{Effects of multiple scattering on angle-independent structural
  color in disordered colloidal materials}

\author{Victoria Hwang}
\thanks{These authors contributed equally to this work. }
\author{Anna B. Stephenson}
  \thanks{These authors contributed equally to this work. }
\affiliation{Harvard John A. Paulson School of Engineering and Applied Sciences, Harvard University, 29 Oxford Street, Cambridge, Massachusetts 02138, USA}
\author{Sofia Magkiriadou}
\altaffiliation{Present address: Laboratory of Experimental Biophysics,
  Institute of Physics, \'{E}cole Polytechnique F\'{e}d\'{e}rale de Lausanne, CH-1015 Lausanne, Switzerland.}
\affiliation{Department of Physics, Harvard University, 17 Oxford Street, Cambridge, Massachusetts 02138, USA}
\author{Jin-Gyu Park}
\altaffiliation{Present address: E Ink Corporation, 1000 Technology Park Drive, Billerica, MA 01821, USA.}
\affiliation{Harvard John A. Paulson School of Engineering and Applied Sciences,
  Harvard University, 29 Oxford Street, Cambridge, Massachusetts 02138,
  USA}
\author{Vinothan N. Manoharan}
 \email{vnm@seas.harvard.edu}
\affiliation{Harvard John A. Paulson School of Engineering and Applied Sciences, Harvard University, 29 Oxford Street, Cambridge, Massachusetts 02138, USA}
\affiliation{Department of Physics, Harvard University, 17 Oxford Street, Cambridge, Massachusetts 02138, USA}

\date{\today}% It is always \today, today,

\begin{abstract}
  Disordered packings of colloidal spheres show angle-independent
  structural color when the particles are on the scale of the wavelength
  of visible light. Previous work has shown that the positions of the
  peaks in the reflectance spectra can be predicted accurately from a
  single-scattering model that accounts for the effective refractive
  index of the material. This agreement shows that the main color peak
  arises from short-range correlations between particles. However, the
  single-scattering model does not quantitatively reproduce the observed
  color: the main peak in the reflectance spectrum is much broader and
  the reflectance at low wavelengths is much larger than predicted by
  the model. We use a combination of experiment and theory to understand
  these features. We find that one significant contribution to the
  breadth of the main peak is light that is scattered, totally
  internally reflected from the boundary of the sample, and then
  scattered again. The high reflectance at low wavelengths also results
  from multiple scattering but can be traced to the increase in the
  scattering cross-section of individual particles with decreasing
  wavelength. Both of these effects tend to reduce the saturation of the
  structural color, which limits the use of these materials in
  applications. We show that while the single-scattering model cannot
  reproduce the observed saturations, it can be used to design materials
  in which multiple scattering is suppressed and the color saturated,
  even in the absence of absorbing components.
  
\begin{description}

% PACS items are no longer really used
%\ocis{290.4020 Mie theory; 290.4210 Multiple scattering; 290.5850 Scattering, particles; 290.5855 Scattering, polarization; 350.4238 Nanophotonics and photonic crystals}

\item[DOI]
\end{description}
\end{abstract}

\maketitle

%%%%%%%%%%%
\section{Introduction}
%%%%%%%%%%%
Structural color comes from constructive interference between waves
scattered from a material with refractive-index variations at the scale
of visible light. When the index variation is periodic, as in photonic
crystals~\cite{john_d._joannopoulos_photonic_2008}, the structural color
is angle-dependent or iridescent. But when the index variation has only
short-range order, the structural color is independent of angle.
Angle-independent structural colors appear matte and homogeneous, often
indistinguishable from colors that come from absorbing pigments. This
type of coloration is found in many species of birds~\cite{prum1998,
  shawkey, dufresne, noh_how_2010, noh_PRE, noh_opt_exp, yin,
  saranathan} and has been mimicked in disordered assemblies of
colloidal particles~\cite{garcia2, takeoka_structural_2009,
  ueno_soft_2009, garcia, harun-ur-rashid_angle-independent_2010,
  forster, lee_quasi-amorphous_2010, retsch, takeoka2, park,
  magkiriadou, yoshioka_production_2014, ge_angle-independent_2014,
  takeoka, teshima_preparation_2015, wang_structural_2016,
  lai_high_2016, yang, iwata, kim_ACS}.

\begin{figure} [htbp]
\centering\includegraphics{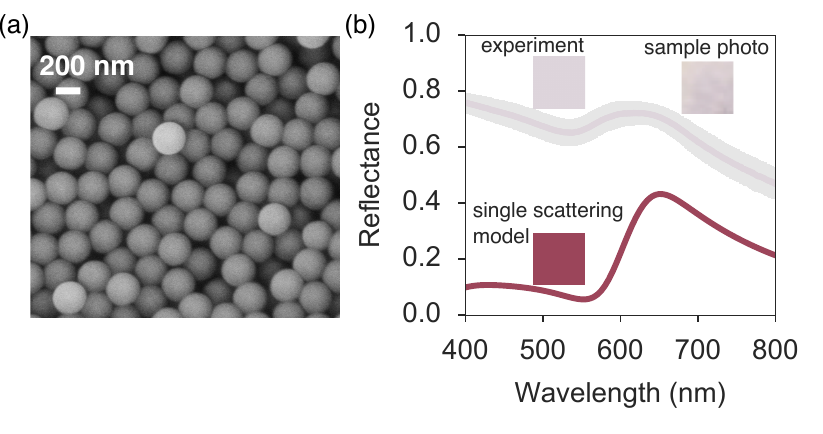}
\caption{(a) Scanning electron micrograph of an angle-independent
  structurally colored film made from \SI{280}{nm} polystyrene spheres.
  (b) Experimental reflectance spectrum (light pink line) and
  predictions of single-scattering model (dark red line) for a
  disordered film of \SI{280}{nm} polystyrene spheres and an effective
  refractive index ranging from 1.322 at \SI{400}{nm} to 1.296 at
  \SI{800}{nm}. Reflectance is measured from all scattering angles with
  an integrating sphere. Line colors are calculated from spectra as
  described in Section~\ref{materials_and_methods}. Error bars are shown
  in grey for each data point and are twice the standard deviation of 6
  measurements from different areas of the sample film. Insets above
  lines are color swatches of the calculated colors. Inset in top right
  is a photograph of the sample.} \label{disordered_packing}
\end{figure}

To explain and predict angle-independent structural color in these
colloidal systems, Magkiriadou and colleagues~\cite{magkiriadou}
developed a model based on a single-scattering approximation and
effective-medium theory. The model, which assumes that the particles are
packed into a glassy arrangement (Fig.~\ref{disordered_packing}a),
predicts that the primary peak in the reflectance spectrum is determined
by the peak of the structure factor, which accounts for constructive
interference arising from the short-range correlations between
particles. The model also predicts that the reflectance spectrum should
include contributions from the form factor, which accounts for the
wavelength-dependent scattering from individual particles. The positions
of the peaks predicted for both the structure and form factors agree
well with those observed in experiment. In comparison to numerical
methods for predicting the color, such as finite-difference time domain
and finite-element methods~\cite{dong, yin, lo, cheng, galinski, xiao,
  chandler}, the single-scattering model gives physical insight into the
reflectance peak and how it varies with the particle size, packing
density, and refractive index.

However, the single-scattering model does not reproduce other features
of the observed reflectance spectra: it underestimates the reflectance
at short wavelengths and the breadth of the primary peak
(Fig.~\ref{disordered_packing}b). These features likely arise from
multiple scattering. Evidence for multiple scattering comes from studies
showing that adding absorbers and reducing the sample thickness
generally increases the saturation of the structural
color~\cite{forster, iwata, yang, takeoka2, cho, retsch}. Also, Noh and
colleagues confirmed that multiple scattering is present in structural
colors produced by sphere-type disordered structures in the barbs of
bird feathers~\cite{noh_PRE}.

Through a combination of experiment and theory, we explain several
multiple scattering effects in the reflectance spectra of disordered
packings of spherical particles. We perform polarization experiments and
reflectance measurements to show that a secondary peak from multiple
scattering explains the breadth of the main color peak. We also use
single-scattering theory to understand the onset of multiple scattering
and its increase at short wavelengths. We validate this physical picture
by developing a design rule that allows one to reduce the amount of
multiple scattering and hence increase the color saturation.

The effects of multiple scattering on backscattering from disordered
colloidal samples have been studied extensively in other contexts, such
as coherent backscattering and Anderson
localization~\cite{wiersma_coherent_1995, froufe-perez_band_2017,
  aubry_resonant_2017, schertel_tunable_2019}, but not nearly to the
same extent in the context of angle-independent structural color.
Therefore, in our study we aim to show how the physical parameters of
the samples---including the particle size and sample thickness---affect
the multiple scattering and hence the color. We anticipate that these
results will be useful in the development of more precise models of
angle-independent structural color and in the formulation of
structurally colored materials.

%%%%%%%%%%%%%%%%
\section{Materials and Methods} \label{materials_and_methods}
%%%%%%%%%%%%%%%%

%_______________________________________________
\subsection{Synthesis of polystyrene particles}
%_______________________________________________

\begin{figure} [htbp]
\centering\includegraphics{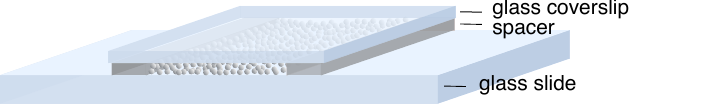}
\caption{Diagram of a sample chamber containing a disordered packing of
  colloidal spheres. The sample chamber consists of a glass slide on the
  bottom, a glass coverslip on top, and Mylar spacers to set the
  thickness. It is sealed with UV-curable epoxy. \label{sample_films}}
\end{figure}

Our structurally colored materials are made from polystyrene particles.
We use emulsion polymerization to synthesize these particles in three
sizes: \SI{280}{nm}, \SI{240}{nm}, and \SI{190}{nm} in diameter. All
materials are used as received. The polymerization reactor consists of a
\SI{500}{\mL} three-necked round-bottom flask equipped with a reflux
condenser, a nitrogen inlet and a mechanical stirrer. In a typical
experiment, we dissolve sodium lauryl sulfate (SLS, 99\%, Aldrich) and
\SI{3.75}{g} of \textit{N}-isopropylacrylamide (NiPAm, 97\%, Aldrich) in
\SI{242.5}{\mL} of deionized (DI) water obtained from a Millipore
Milli-Q system in the reactor. We control the diameter of the
polystyrene spheres through the amount of SLS. We use \SI{95}{mg} SLS
for the \SI{280}{nm} particles, \SI{190}{mg} for \SI{240}{nm}, and
\SI{285}{mg} for \SI{190}{nm}. We add the NiPAm so that we can further
functionalize the particles for other experiments not described in this
paper~\cite{park_photoniccrystal_2017}. We then add \SI{71.25}{g} of
styrene (99\%, Aldrich) under vigorous stirring. We heat the mixture to
\SI{80}{\celsius} and add \SI{180}{mg} of potassium persulfate (KPS,
99\%, Aldrich) dissolved in \SI{7.5}{mL} of DI water. The reaction runs
for \SI{8}{\hour}. Finally, we wash the resulting particles by dialysis
against DI water for \SI{5}{\day}. We measure the particle diameters and
polydispersity by image analysis of scanning electron micrographs (SEM).
Because the polydispersity index is only 2\%, we assume that the
particles are monodisperse in most of our scattering calculations (see
section~\ref{sec:single-scatt-model}).

%_______________________________________________
\subsection{Sample preparation for polarization experiments}
%_______________________________________________

For our polarization experiments, we make disordered, structurally
colored films from the polystyrene particles described above
(polydispersity 2\%). We start by centrifuging the particles in
\SI{25}{mM} NaCl for \SI{30}{\minute} at \SI{14000}{\g} and removing the
supernatant. The salt screens the electrostatic interactions between the
particles, which is sufficient to prevent them from crystallizing. We
then vortex the mixture for \SI{5}{\minute} to resuspend the particles
in the remaining liquid. The concentration of the resulting suspension
is 45\% w/w in water.

We make the films by drying these suspensions in sample chambers of
controlled thicknesses, which we make from Mylar spacers sandwiched
between glass slides and glass coverslips (Fig.~\ref{sample_films}).
Each chamber is sealed with UV-curable epoxy (Norland Optical Adhesive
68). We pipette a dense suspension into a sample chamber of thickness
\SI{77}{\um} and leave a pool of excess suspension at the inlet of the
sample chamber. As the water evaporates from the opposite end, more
suspension is pulled into the chamber from the pool through capillary
action. We periodically replenish the pool as its volume is pulled into
the sample chamber. Over the course of \SIrange{6}{8}{\hour}, the
particles become densely packed as the water evaporates. We dry the
films overnight at room temperature and remove any excess water by
drying them in an oven at \SI{60}{\celsius} for several hours. We then
seal the sample chamber with 5-minute epoxy (No. 14250, Devcon). The
film of \SI{280}{nm} polystyrene particles measured in this paper has an
area of \SI{2.8}{\cm} $\times$ \SI{1.9}{\cm} and a thickness of
\SI{77}{\um}. We estimate the volume fraction from the weight of the
polystyrene after drying, the density of the polystyrene, and the volume
of the sample chamber.

%_______________________________
\subsection{Polarization measurements}
%_______________________________

\begin{figure} [htbp]
\centering\includegraphics{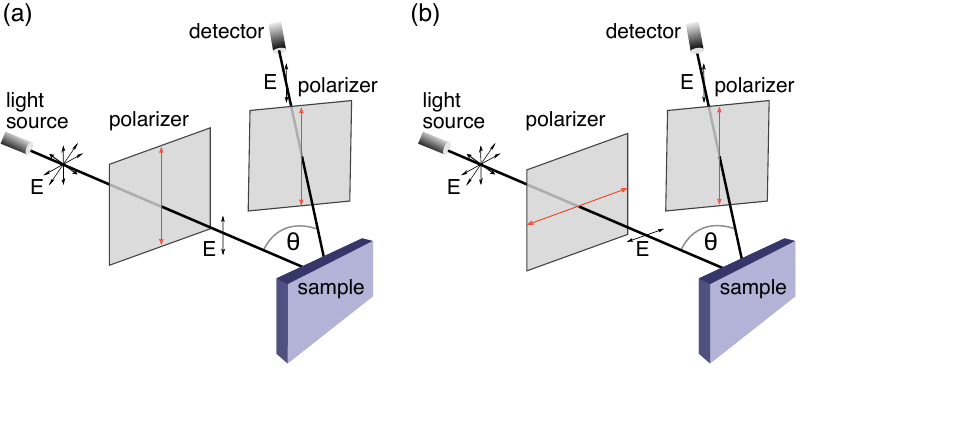}
\caption{Setup for polarization measurements. Light from a monochromator
  shines through a linear polarizer and onto the sample. The light that
  scatters at an angle $\theta$ to the normal is detected through either
  (a) a parallel or (b) a perpendicular polarizer. Red arrows indicate
  the axis of transmission of the polarizer. \label{goniometer}}
\end{figure}

We determine the wavelength-dependence of multiple scattering by
measuring the spectrum of polarized light. The entire setup is contained
inside a spectrophotometer (Agilent Cary 7000 Universal Measurement
Spectrophotometer). The sample is mounted on a goniometer (Universal
Measurement Accessory of the Agilent Cary 7000 Universal Measurement
Spectrophotometer). We set the detection angle $\theta$ to $\ang{16}$,
the smallest angle for which the detector does not cross the incident
beam. We choose the largest available aperture, which subtends an angle
of $\ang{12}$, defined by the two edges of the detector and the center
of the setup. We also performed measurements with detection angles
$\theta=\ang{22}$--$\ang{76}$ to characterize the angle-dependence
of our samples (see Fig.~\ref{angle-dependence}).

We illuminate the sample with light from a monochromatic source (double
out-of-plane Littrow monochromator) sent through a linear polarizer, as
shown in Fig.~\ref{goniometer}. The illuminated spot on the sample is a
\SI{5}{\mm} $\times$ \SI{5}{\mm} square. We detect the scattered light
through a second polarizer placed in front of the detector (R928
Hamamatsu photomultiplier tube). We measure at wavelength intervals of
\SI{1}{\nm} for \SI{0.3}{\s} at each wavelength. To measure the diffuse
reflectance, we use the same source and spectrophotometer
(Fig.~\ref{disordered_packing}), but we remove the polarizers and use an
integrating sphere accessory instead of the Universal Measurement
Accessory. In the integrating sphere measurements, the illuminated spot
on the sample is \SI{1}{\mm} $\times$ \SI{3}{\mm}, and we measure at
wavelength intervals of \SI{1}{\nm} for \SI{0.1}{\s} at each wavelength.

We measure both the co-polarized spectrum, which includes any singly
scattered light as well as any multiply scattered light that returns to
its initial polarization, and the cross-polarized spectrum, which
includes only multiply scattered light. In our measurements, light is
incident on the glass slide of the sample chamber. We normalize these
spectra to correct for the reflection from the glass slide and for the
wavelength-dependence of the incident beam and the polarizers:
\begin{equation}
R_{\textnormal{co}} = \frac{I_{\textnormal{sample,co}}-I_{\textnormal{glass,co}}}{I_{\textnormal{inc}}T_{\textnormal{1v}}T_2}
\label{eq:co}
\end{equation}
and
\begin{equation}
R_{\textnormal{cr}} = \frac{I_{\textnormal{sample,cr}}-I_{\textnormal{glass,cr}}}{I_{\textnormal{inc}}
          T_{\textnormal{1h}}T_2}, 
\label{eq:cross}
\end{equation}
where $I_{\textnormal{sample, co/cr}}$ is the intensity of light
scattered from the sample in the co/cross-polarized setup,
$I_{\textnormal{glass, co/cr}}$ is the intensity of light scattered from
a glass slide in the co/cross-polarized setup, and
$I_{\textnormal{inc}}$ is the source intensity. $T_{\textnormal{1v/1h}}$
and $T_2$ are the measured transmittances of the two polarizers:
\begin{equation}
\begin{split}
T_{\textnormal{1v}} = \frac{I_{\textnormal{1v,out}}}{I_\textnormal{inc}}\\
T_{\textnormal{1h}} = \frac{I_{\textnormal{1h,out}}}{I_\textnormal{inc}}\\
T_2 = \frac{I_\textnormal{2v,out}}{I_{\textnormal{1v,out}}},
\end{split}
\end{equation}
where $I_{\textnormal{1v,out}}$ is the intensity of light measured
through the first polarizer in the co-polarized setup and
$I_{\textnormal{1h,out}}$ is the intensity of light measured through the
first polarizer in the cross-polarized setup. $I_{\textnormal{2v,out}}$
is the intensity measured through two polarizers oriented vertically. To
account for sample inhomogeneity, we report the mean of the reflectance
of five separate spots on the sample.

We quantify the amount of multiple scattering through the depolarization
ratio \cite{noh_PRE}
\begin{equation}
D(\lambda) = \frac{R_{\textnormal{cr}}(\lambda)}{R_{\textnormal{co}}(\lambda)},
\label{eq:depolarization}
\end{equation}
where $R$ is reflectance. Pure, high-order multiple scattering should
lead to a depolarization of unity, while pure single scattering should
lead to a depolarization of zero, because all of the light retains its
initial polarization.

We can estimate the amount of single and multiple scattering in the
sample by assuming that any light scattered more than once is randomly
polarized. Because the polarization of low-order multiple scattering may
not be completely randomized, this is a coarse approximation, but it
provides a useful estimate. Under this approximation, half of the
multiply scattered light is detected through crossed polarizers and the
other half is detected through parallel polarizers, while all of the
singly scattered light is detected through parallel polarizers. Our
estimate of the multiply scattered signal is therefore
\begin{equation}
  R_{\textnormal{multiple scat}} = 2 R_{\textnormal{cr}},
\end{equation}
and our estimate of the singly scattered signal is
\begin{equation}
  \label{eq:single_scat_extracted}
  R_{\textnormal{single scat}} = R_{\textnormal{co}} - R_{\textnormal{cr}}.
\end{equation}

% ______________________________________________
\subsection{Sample preparation and measurements of thickness-controlled films}
\begin{figure}[htbp]
  \centering\includegraphics{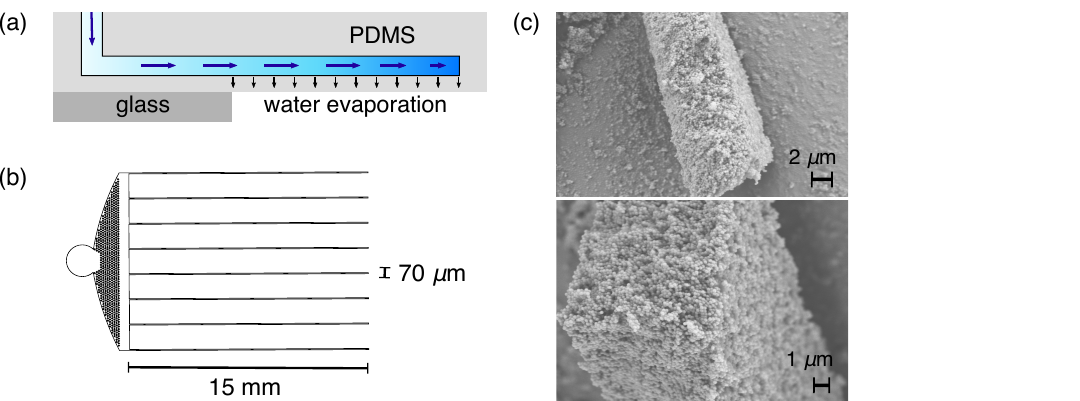}
  \caption{(a) Microevaporator design \cite{merlin}. Blue arrows
    indicate direction of particle flow, and blue color indicates
    particle packing density. (b) Mask design for the microevaporators.
    (c) SEM micrographs of assembled structural color
    films. \label{fig:microevaporators_design}}
\end{figure}

%_______________________________________________
To determine how the structural color varies with sample thickness, we
build polydimethylsiloxane (PDMS) microevaporators with controlled
thickness, following the microfluidic protocol in Ref.~\citenum{merlin}.
We make microevaporator channels that are \SI{70}{\um} wide and have
thicknesses of \SIlist{7; 19; 33; 47}{\um}
(Fig.~\ref{fig:microevaporators_design}). We then inject a binary
suspension of polystyrene particles (by volume, 2 parts of
\num{240}-\si{nm}-diameter and 1 part of \num{190}-\si{nm}-diameter,
each at 0.5\% v/v) and \SI{50}{mM} of sodium chloride into the channels,
which have an inlet port but no outlet. As the water evaporates through
a \num{15}-\si{\um} layer of PDMS on the bottom of the channel, the
particles pack into a film. Because the evaporation of water is slow
(\SIrange{1}{3}{\day} at room temperature), adding salt is not
sufficient to prevent crystallization. Therefore we use binary
suspensions in addition to salt. From the particle size and the location
of the main color peak, we estimate the resulting volume fraction as
0.5, using our single-scattering model. We use this approach instead of
the gravimetric estimate we use for the polarization experiments because
the microevaporated films do not pack densely along the entirety of
their \num{15}-\si{mm} length. We measure reflection spectra only in the
densely packed regions.

We measure the reflectance spectra of the films with a fiber-optic
spectrometer (Ocean Optics HR2000+) attached to an optical microscope
(Nikon Eclipse LV-100). We illuminate the films with collimated white
light from a halogen lamp, and we collect the scattered light with a
50$\times$ objective (Nikon LU Plan Fluor, NA = 0.8). We normalize the
reflection data against the reflection spectrum of an aluminum mirror.

To quantify the differences in color between the films, we calculate the
color saturation using the CIELUV coordinates, which form a perceptual
color space:
\begin{equation}\label{eq:saturation}
s_{uv} = \frac{C^*_{uv}}{L^*} = \frac{\sqrt{(u^*)^2 + (v^*)^{2} }}{L^*},
\end{equation}
where $C^*_{uv}$ is the chroma, $L^*$ corresponds to lightness, and
$u^*$ and $v^*$ correspond to chromaticity~\cite{iwata}. To obtain the
($L^*$, $u^*$, $v^*$) values, we first calculate the ($X$, $Y$, $Z$)
color values by integrating the intensity spectrum from the reflectance
data multiplied by matching functions that account for the average
chromatic response of the human eye \cite{cieluv, cieluv2, cieluv3}.
Then we calculate the ($L^*$, $u^*$, $v^*$) values using the following
transformation:
\begin{align}
L^* &= \begin{cases}
  \left(\frac{29}{3}\right)^3 Y / Y_n,&   Y / Y_n \le \left(\frac{6}{29}\right)^3 \\
  116 \left( Y / Y_n \right)^{1/3} - 16,&  Y / Y_n  >   \left(\frac{6}{29}\right)^3      
\end{cases}\\
u^* &= 13 L^* (u^\prime - u_n^\prime) \\
v^* &= 13 L^* (v^\prime - v_n^\prime),
\end{align}
where $u^\prime$ and $v^\prime$ are calculated from the (X, Y, Z) color values:
\begin{align}
u^\prime &= \frac{4 X}{X + 15 Y + 3 Z}  \\
v^\prime &= \frac{9 Y}{X + 15 Y + 3 Z}.
\end{align}
and where $u_n^\prime$ and $v_n^\prime$ are calculated using the above
equations, with the ($X$, $Y$, $Z$) color values of a perfect diffuse
reflector, ($X_n$, $Y_n$, $Z_n$), as defined for the CIE Standard
Illuminant D65. We use the software package ColorPy~\cite{kness} to
perform this calculation.

% ______________________________________________
\subsection{Single-scattering model calculations}
% _______________________________________________
\label{sec:single-scatt-model}

For our calculations, we use the single-scattering model of Magkiriadou
and colleagues~\cite{magkiriadou}, but we use the Bruggeman formula for
the effective refractive index of the sample
\cite{bruggeman_berechnung_1935, bruggeman_berechnung_1936,
  markel_introduction_2016} instead of the Maxwell-Garnett
approximation. The Bruggeman formula is symmetric and should therefore
be more reliable than Maxwell-Garnett at volume fractions near
0.5~\cite{markel_introduction_2016}, like those in our samples. We
account for dispersion in the materials by using the Sellmeier
dispersion formula for polystyrene, with parameters that are fit to
experimental data~\cite{sultanova_dispersion_2009}. Thus, the effective
index also varies with wavelength, as we report in the figure captions
describing our measurements.

As discussed in Ref.~\citenum{magkiriadou}, the model accounts not only
for interference between waves scattered from different particles, but
also for interference effects within the particles, which lead to
backscattering resonances at certain wavelengths. We do, however,
neglect near-field effects that might occur in dense packings
\cite{aubry_resonant_2017, schertel_tunable_2019,
  rezvani_naraghi_near-field_2015}. Such effects become important when
the transport length (see Section~\ref{sec:results-discussion}) is
comparable to the wavelength. We estimate the transport length from the
energy-density coherent-potential approximation
(ECPA)~\cite{aubry_resonant_2017}, a more sophisticated approximation
that accounts for near-field effects. According to Fig.~4 of
Ref.~\citenum{aubry_resonant_2017}, for our disordered samples, which
have a volume fraction of 0.5 and particle-radius-to-wavelength ratios
that are less than 0.4, the transport length calculated by ECPA at
resonance is eight times the wavelength, and much larger off resonance.
Thus, for the small particle sizes (relative to the wavelength) that we
use in our samples, it is reasonable to neglect the near-field effects
over most of the wavelengths in our measurements. Our approximation is
further justified by the good agreement between model and experiment for
the position of the main color peak (Fig.~\ref{single_multiple_scat}).

We also modify the model to account for the two particle sizes used in
our microevaporator films. Following Scheffold and
Mason~\cite{scheffold_scattering_2009}, we write the scattered intensity
$I$ as
 \begin{equation}
   I \propto \overline{F_M(q)} S_M(q)
\end{equation}
where $\overline{F_M(q)}$ is the polydisperse form factor (see below)
and $S_M(q)$ is the polydisperse structure factor derived by Ginoza and
Yasutomi \cite{ginoza_measurable_1999}. This structure factor assumes a
Schulz size distribution, which tends to a Gaussian distribution when
the polydispersity is small. Scheffold and Mason found good agreement
between this structure factor and experimental measurements. We calculate the polydisperse form factor $\overline{F_M(q)}$ from a
size-average:
\begin{equation}
  \overline{F_M(q)} = \int_0^\infty f(\sigma) F(q \sigma) \, d\sigma,
\end{equation}
where $f(\sigma)$ is the Schulz distribution,
\begin{equation}
  f(\sigma) = \left(\frac{t+1}{\sigma_0}\right)^{t+1} \, \frac{\sigma^t}{t!} \, \exp{\left(-\sigma \, \frac{t+1}{\sigma_0}\right)},
\end{equation}
$F(q\sigma)$ is the monodisperse form factor from Mie theory, $\sigma$
the particle diameter, and $\sigma_0$ the mean of the size distribution.
The parameter $t= (1-p^2)/p^2$, where $p$ is the polydispersity index,
which accounts for the width of the distribution. For the binary samples
used in the microevaporator measurements, we calculate the binary
polydisperse form factor as the weighted average of the individual
polydisperse form factors.

%%%%%%%%%%%%%%%%%
\section{Results and Discussion}
\label{sec:results-discussion}
%%%%%%%%%%%%%%%%%
\begin{figure}[htbp]
  \centering\includegraphics{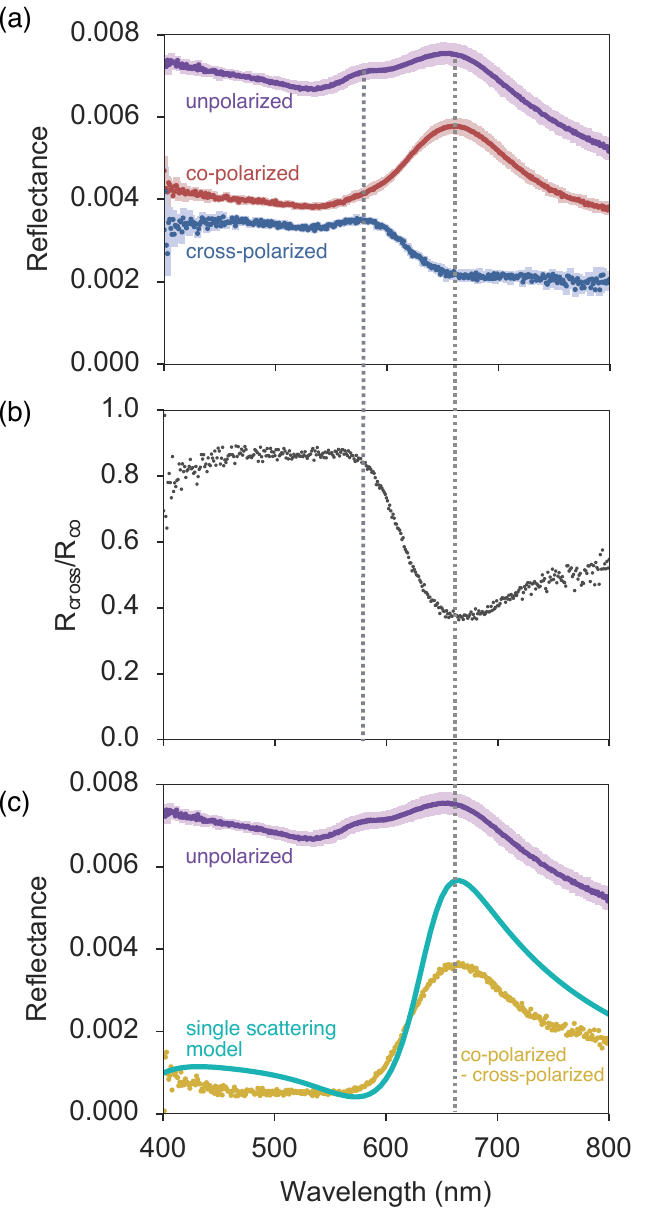}
  \caption{(a) Unpolarized (purple), co-polarized (red), and
    cross-polarized (blue) reflectance spectra of a disordered packing
    of \num{280}-\si{nm} polystyrene spheres. Error bars are shown by
    the shaded regions around each measured spectrum and are twice the
    standard deviation of five measurements from different areas of the
    sample film. (b) Depolarization ratio, calculated from
    Eq.~\eqref{eq:depolarization}. (c) Single-scattering reflectance
    spectrum (yellow) extracted from data using
    Eq.~\eqref{eq:single_scat_extracted}, compared to spectrum (cyan)
    calculated from our single-scattering model
    (Section~\ref{sec:single-scatt-model}). The dashed lines indicate
    the locations of the primary (\SI{660}{nm}) and secondary
    (\SI{580}{nm}) peaks. The calculation uses the following parameters:
    a volume fraction of 0.53, corresponding to the measured volume
    fraction, a diameter of \SI{280}{nm}, and an effective refractive
    index ranging from 1.322 at \SI{400}{nm} to 1.296 at \SI{800}{nm}.
    See Fig.~\ref{angle-dependence} for the angle-dependence of the co-
    and cross- polarized spectra.}
  \label{single_multiple_scat}
\end{figure}

We observe three distinct features in the reflectance spectrum of a
\num{77}-\si{\um} film made from a disordered packing of
\num{280}-\si{nm} polystyrene spheres: a primary peak near \SI{660}{nm},
a secondary peak near \SI{580}{nm}, and an increase in reflectance with
decreasing wavelength (Fig.~\ref{single_multiple_scat}a). We examine
this sample in detail because its spectral features are well separated
by wavelength, which allows us to study each feature independently. The
secondary peak is visible in measurements over a narrow range of
detection angles (as in Fig.~\ref{single_multiple_scat}a) but not in a
measurement using an integrating sphere, where the peaks are broadened
and merged (as in Fig.~\ref{disordered_packing}b).

Both the polarization experiments and the single-scattering model
suggest that the measured primary peak near \SI{660}{nm} comes from
single scattering: the depolarization ratio is minimized at the peak
wavelength (Fig.~\ref{single_multiple_scat}b), and the model predicts a
peak at the same wavelength. The absence of this peak in the
cross-polarized spectrum suggests that multiple scattering does not
contribute to the peak. The agreement with the single-scattering model
suggests that the primary peak is due to the interference between waves
scattered from the structure.

Indeed, we find that the predictions of the single-scattering model
agree well with the single-scattering spectrum extracted from the data
through Eq.~\eqref{eq:single_scat_extracted}, as shown in
Fig.~\ref{single_multiple_scat}c. The subtraction of the cross-polarized
spectrum removes much of the low-wavelength intensity and narrows the
primary peak, resulting in a spectrum that more closely matches that
predicted by the single-scattering model. This agreement tells us that
the single-scattering model not only predicts the primary peak of the
reflectance, but also gives a reasonable estimate for how much single
scattering contributes to the reflectance.

\begin{figure}[htbp]
  \centering\includegraphics{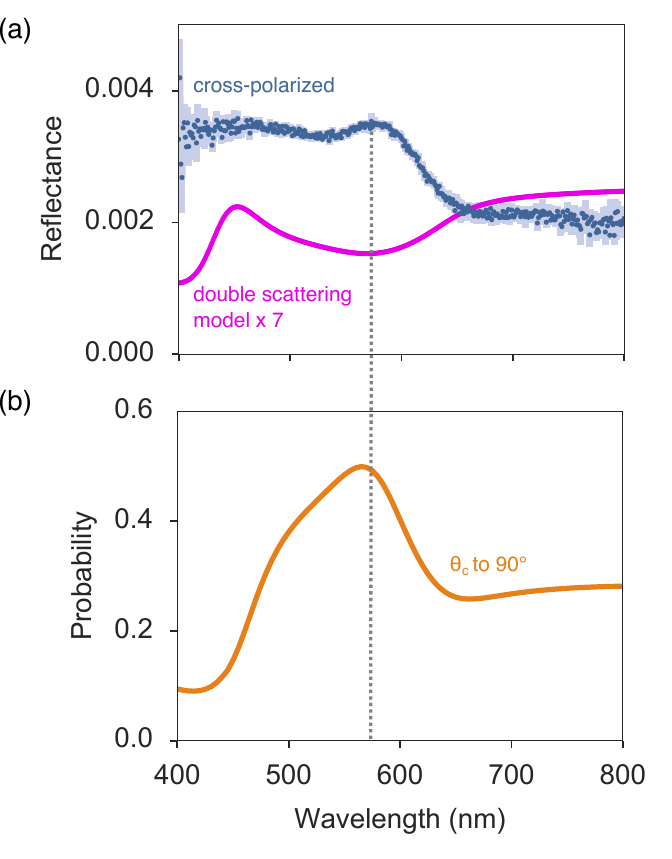}
  \caption{Total internal reflection, rather than double scattering,
    explains the peak in the cross-polarized spectrum. (a)
    Cross-polarized (blue) spectrum from
    Fig.~\ref{single_multiple_scat}a and reflectance calculated with a
    double-scattering model (magenta) of a disordered packing of
    \num{280}-\si{nm} polystyrene spheres, assuming a volume fraction of
    0.53 and detection angles ranging from $\theta=$ \ang{10} to
    \ang{22} (see Fig.~\ref{goniometer}). The effective refractive
    index, calculated using the Bruggeman approximation, ranges from
    1.322 at \SI{400}{nm} to 1.296 at \SI{800}{nm}. The
    double-scattering reflectance is multiplied by a factor of 7 for
    clarity. Error bars for the cross-polarized spectrum are shown in
    light blue and are twice the standard deviation of 5 measurements
    from different areas of the sample film. (b) Probability of single
    scattering into the totally internally reflected angular range. The
    critical angle $\theta_c$ for this sample, as calculated from the
    wavelength-dependent effective refractive index, ranges from
    \ang{49} at \SI{400}{nm} to \ang{50} at \SI{800}{nm},}
\label{double_scattering_TIR}
\end{figure}

Having explained the origin of the primary peak in terms of single
scattering, we now turn to the spectral features that deviate from the
predictions of the model: the secondary peak and the increase in
reflectance toward low wavelengths. The secondary peak in the
unpolarized spectrum comes from the peak in the cross-polarized
reflectance, which is due to multiple scattering. Following the example
of Noh and coworkers~\cite{noh_PRE}, we first examine whether double
scattering can explain this peak.

In our double-scattering model, the reflectance is proportional to the
integral of the phase function of two consecutive scattering events. The
phase function is the probability that light is scattered in a certain
direction $\theta'$:
\begin{equation}\label{phase_function}
p(\theta') = \frac{1}{\sigma_{\mathrm{scat}}} \frac{\mathrm{d}\sigma_{\mathrm{scat}}}{\mathrm{d}\Omega} (\theta'),
\end{equation}
where the scattering angle $\theta' = 180^{\circ} - \theta$, $
\mathrm{d}\sigma_{\mathrm{scat}}/\mathrm{d}\Omega$ is the differential
scattering cross-section, and $\sigma_{\mathrm{scat}}$ is the total
scattering cross-section~\cite{bohren}. We assume that in our disordered
samples, the phase function is isotropic in the azimuthal angle $\phi$.
The double-scattering phase function is
\begin{equation}
p_{\mathrm{double}} = p_{\mathrm{first}} p_{\mathrm{second}},
\end{equation}
where $p_{\mathrm{first}}$ and $p_{\mathrm{second}}$ are the phase
functions of the first and second scattering events. Both are calculated
from the single-scattering model, $p_{\mathrm{second}}$ by rotating
$p_{\mathrm{first}}$ from the lab frame to the scattering plane of the
second event.

Double scattering does not explain the secondary peak in our samples, as
shown in Fig.~\ref{double_scattering_TIR}a. Noh and colleagues performed
double-scattering calculations based on small-angle X-ray scattering
data and found that double scattering does appear to explain the
secondary peak in the reflectance of cotinga feathers~\cite{noh_PRE}.
Our results may differ from theirs because of differences in structure:
our structures consist of polystyrene spheres in an air matrix, whereas
bird feathers are ``inverse'' structures of air spheres in a keratin
matrix.

In our samples, the secondary peak appears to be due to scattering from
totally internally reflected waves. When light is scattered toward the
sample interface, some fraction is totally internally reflected back
into the sample, where it can scatter again. To contribute to the
reflectance, the totally internally reflected light must scatter at
least once more before it exits the sample. By integrating the phase
function for a single-scattering event over the angular range for total
internal reflection, we find the probability that singly scattered light
is totally internally reflected. In this calculation, we consider the
interface to be between sample and air, since the refraction due to the
glass slide cancels when we apply Snell's law to both the sample-glass
and glass-air boundaries.

The probability is peaked at a wavelength that matches that of the
observed secondary peak (Fig.~\ref{double_scattering_TIR}b). Although
the probability does not quantitatively predict the contribution of
totally internally reflected light to the reflection spectrum---a more
sophisticated model is needed for such a prediction---the agreement
between the peak positions strongly suggests that the secondary peak is
due to totally internally reflected light. The single-scattering model
does not capture this effect, because the model assumes that totally
internally reflected light is lost. We note that this effect might also
be present in the bird feathers examined by Noh and colleagues.

The effect of internal reflection on multiply scattering media has been
investigated by Zhu and colleagues \cite{zhu_internal_1991}, but their
model assumes that light from inside the sample impinges on the boundary
over a uniform distribution of angles, which is not the case for the
single and low-order multiple scattering in our samples. The
angle-dependence of singly scattered light in our samples comes from the
angle-dependence of the structure factor and the form factor, which are
used to calculate the differential scattering cross-section.

\begin{figure}[htbp]
  \centering\includegraphics{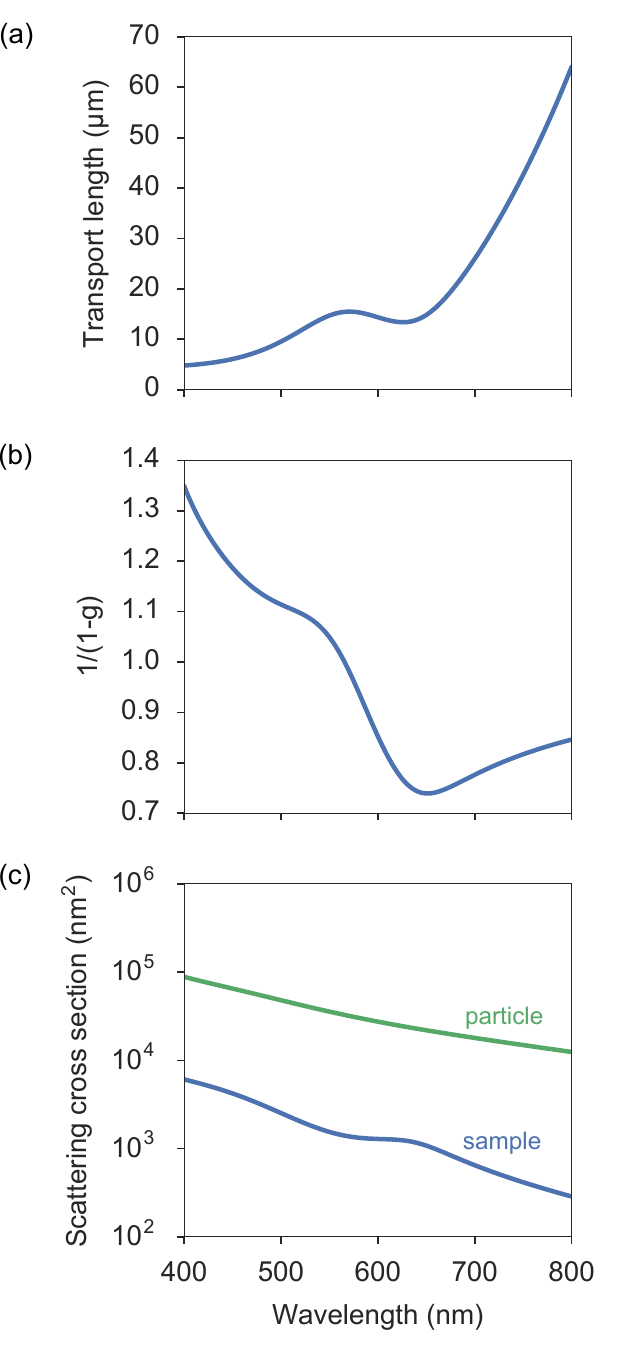}
  \caption{(a) Transport length, (b) asymmetry parameter factor, and (c)
    scattering cross-sections calculated for a sample with
    \num{280}-\si{nm} polystyrene spheres at a volume fraction of 0.53
    and an effective refractive index ranging from 1.322 at \SI{400}{nm}
    to 1.296 at \SI{800}{nm}. The particle cross-section is calculated
    with Mie theory, and the sample cross-section is calculated with the
    single-scattering model. Note that the cross section is on a log
    scale. \label{fig:transport_length_cross_section_g_contribution}}
\end{figure}

\begin{figure*}[htbp]
  \centering\includegraphics{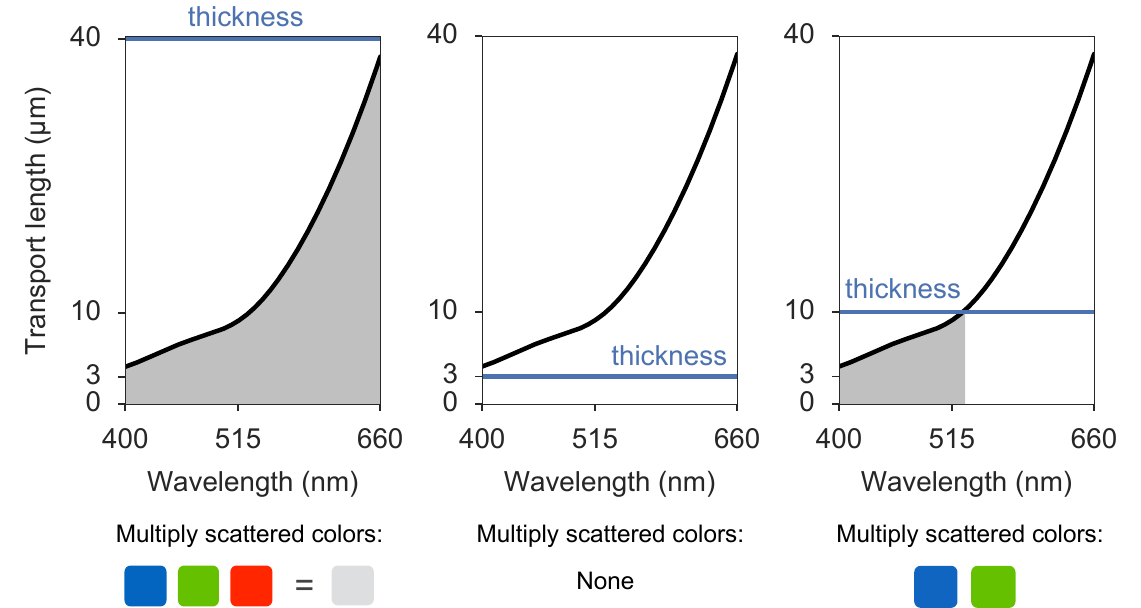}
  \caption{Transport length calculations for a disordered packing of a
    binary mixture of polystyrene particles (2 parts by volume of
    \num{240}-\si{nm}-diameter and 1 part of \num{190}-\si{nm}-diameter, each with a polydispersity of 2\%)
    in air assuming a volume fraction of 0.5 and an effective refractive
    index ranging from 1.302 at \SI{400}{nm} to 1.285 at \SI{660}{nm}.
    The panels show schematically how changing the sample thickness
    relative to the transport length changes how certain colors are
    reflected from the sample.\label{fig:transport_length}}
\end{figure*}

To understand the origin of the large depolarization at low wavelengths
and the accompanying rise in scattering toward the blue, we return to
the single-scattering model. Although this model cannot capture the
contribution of multiple scattering to the spectrum, it can predict the
propensity for multiple scattering. This propensity is characterized by
the transport length $l^*(\lambda)$, which is the distance that light
propagates into the sample before its direction is
randomized~\cite{kaplan_diffuse-transmission_1994,
  rojas-ochoa_photonic_2004, reufer_transport_2007}. The shorter the
transport length at a fixed film thickness, the higher the propensity
for multiple scattering.

The transport length $l^*$ is related to the asymmetry parameter $g$ and
scattering length $l_{\mathrm{scat}}$:
\begin{equation}\label{eq:transport_length}
l^* = \frac{l_{\mathrm{scat}}}{1-g},
\end{equation}
where $g=\langle \cos(\theta) \rangle$ and $\theta$ is the scattering
angle. The scattering length $l_{\mathrm{scat}}$ is the average distance
between scattering events and is calculated as $1/(N
\sigma_{\mathrm{scat}})$, where $N$ is the number density and
$\sigma_{\mathrm{scat}}$ is the scattering cross-section of the sample
\cite{ishimaru}. We can then express the transport length $l^*$ as
\begin{equation}\label{eq:transport_length2}
  l^* = \frac{1}{N\sigma_{\mathrm{scat}}(1-g)}.
\end{equation}
We use our single-scattering model to calculate the sample scattering
cross-section and the asymmetry parameter, which includes the
contributions of both the form factor and the structure factor within
our effective-medium approximation~\cite{magkiriadou}.

Our calculations show that the transport length has a local minimum at
approximately the wavelength of the primary reflectance peak
(Fig.~\ref{fig:transport_length_cross_section_g_contribution}a). This
minimum is not surprising, since constructive interference contributes
to strong backscattering at the structural
resonance~\cite{rojas-ochoa_photonic_2004}, leading to both a minimum in
the asymmetry parameter factor $1/(1-g)$ (see
Fig.~\ref{fig:transport_length_cross_section_g_contribution}b) and a
local maximum in the sample scattering cross-section (see
Fig.~\ref{fig:transport_length_cross_section_g_contribution}c). Although
a minimum in the transport length should correspond to a higher
propensity for multiple scattering, the cross-polarized spectrum does
not show a peak at the same wavelength, indicating that high-order
multiple scattering does not contribute significantly to the main peak.

More interestingly, the transport length for this sample is smallest
(less than \SI{10}{\um}) at wavelengths from \SIrange{400}{500}{nm}, as
shown in Fig.~\ref{fig:transport_length_cross_section_g_contribution}a.
This feature does not appear to be due to the asymmetry parameter
factor, $1/(1-g)$, which increases with decreasing wavelengths. If this
factor were the only contribution to the transport length, the transport
length would also increase with decreasing wavelength, in contradiction
with our results.

Instead, the decrease in transport length at short wavelengths appears
to be due to the scattering from individual particles. We find that the
single-particle scattering cross-section, calculated with Mie theory,
increases by a factor of 7 from \SIrange{800}{400}{nm}, leading to an
increase in the sample cross-section by more than an order of magnitude
at short wavelengths
(Fig.~\ref{fig:transport_length_cross_section_g_contribution}c).

The transport length calculations allow us to explain how the structural
color changes with the thickness of the sample, as illustrated in
Fig.~\ref{fig:transport_length}. As the sample thickness changes in
comparison to the transport length, different colors can appear in the
reflectance spectra. When the thickness is larger than the transport
length in the entire spectrum, all wavelengths are likely to be multiply
scattered, and the resulting color is white. When the thickness is
smaller than the transport length at any wavelength, light is likely to
pass through the film without scattering, and the sample becomes
transparent. When the thickness is approximately the transport length at
the reflectance peak, the resulting structural color is saturated,
meaning that, on average, the scattering at wavelengths close to the
reflectance peak is high relative to that at other wavelengths. Although
the diagram in Fig.~\ref{fig:transport_length} does not allow one to
quantitatively predict the color, it illustrates the important physical
considerations needed to design saturated colors.

\begin{figure}[htbp]
\centering \includegraphics{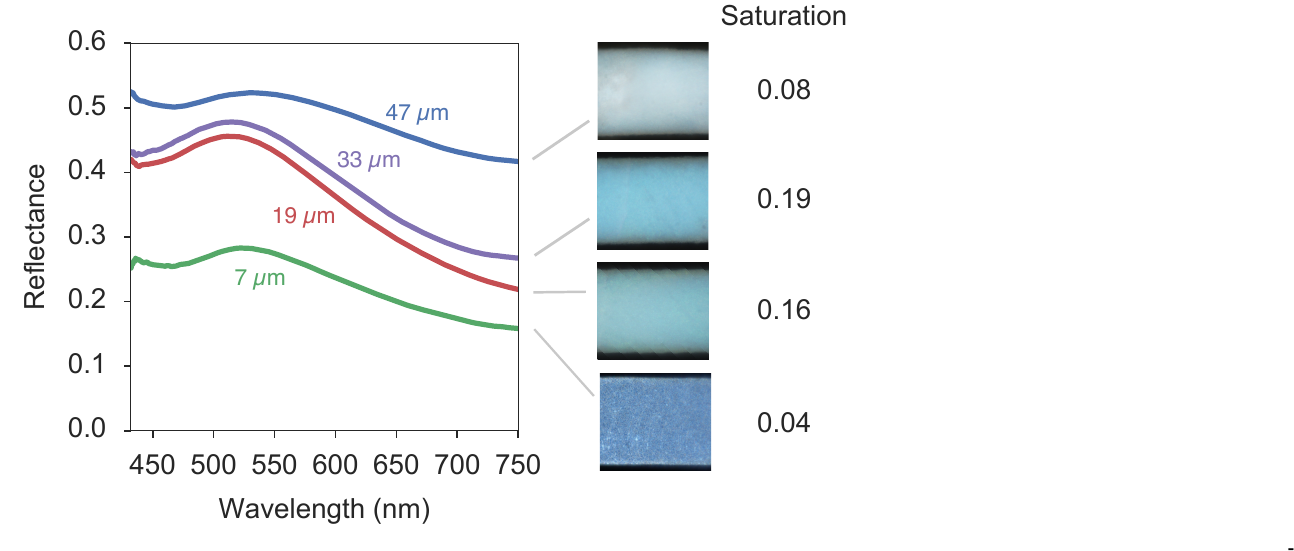}
\caption{Reflectances of films made in microevaporators of different
  thicknesses, as indicated by the labels near each curve. The films
  consist of packings of polystyrene nanoparticles (2 parts of
  \num{240}-\si{nm}-diameter and 1 part of \num{190}-\si{nm}-diameter by
  volume, each with a polydispersity of 2\%). At right, dark-field optical micrographs show the color at
  each thickness. The color saturations are listed to the right of each
  image. \label{fig:microevaporators_results}}
\end{figure}

To validate this physical picture, we assemble four structurally colored
films of different thicknesses inside microevaporators and measure the
reflectance spectrum at each thickness. We use the saturation of the
spectrum as a measure of the amount of multiple scattering in each
sample. The highest color saturations occur at thicknesses of
\SIlist{19; 33}{\um} (Fig.~\ref{fig:microevaporators_results}), which
are 2--3 times larger than the transport length (\SI{10}{\um}; see
Fig.~\ref{fig:transport_length}) calculated at the primary peak. Samples
smaller than this transport length are translucent, as shown by the low
reflectance of the \num{7}-\si{\um} film. In fact, in the image of this
film, the dark background is visible through the sample. When the sample
is thicker than the transport length at all wavelengths, the color is
desaturated, as shown by the spectrum and image of the \num{47}-\si{\um}
film. These results agree qualitatively with the predictions shown in
Fig.~\ref{fig:transport_length}.

In addition, the main peaks in the more saturated samples are narrower
and slightly shifted with respect to the peaks in the thickest and
thinnest films. In the thickest film, scattering at wavelengths off
resonance shifts the location of the peak. In the thinnest film, the
peak is flattened because the sample has less scattering overall. Also,
this film shows a much lower reflection at short wavelengths, consistent
with our analysis in
Figure~\ref{fig:transport_length_cross_section_g_contribution}. The
changes in the shapes of the peaks with changing thickness illustrate
the central role that light transport plays in setting the color.

We conclude that although the single-scattering model cannot
quantitatively predict the reflectance spectrum of samples with large
thicknesses, it can predict how the multiple scattering (and color
saturation) varies with wavelength and sample thickness. This prediction
comes from calculating the transport length as a function of wavelength
and comparing it to the sample thickness. Though the transport length is
commonly used to understand multiple scattering in contexts such as weak
localization~\cite{wiersma_coherent_1995, aubry_resonant_2017,
  schertel_tunable_2019}, our results show that it is also useful for
understanding how multiple scattering compromises color saturation---for
example, through the rise in reflectance at short wavelengths.

The model can be used to design saturated colors in the absence of
absorption: after calculating the transport length as a function of
wavelength, one can make a sample with saturated color by choosing a
film thickness that is on the order of (more precisely, 2 to 3 times)
the transport length at the reflection peak. In the absence of
absorption, this design rule is quantitative, in that the transport
length calculated from the single-scattering model provides a
quantitative estimate for the sample thickness required for optimal
saturation. The estimate is coarse but provides a useful starting point
for sample design.

In the presence of broadband absorbers, which are often used to tune
saturation~\cite{forster, iwata}, the transport length is still an
important parameter. But now a third lengthscale must be taken into
account: the absorption length, which is set by the concentration of
absorbing material, among other factors. We consider several possible
orderings of these three length scales, assuming, for simplicity, that
the absorption length does not vary significantly with wavelength. Let
$l_\textnormal{abs}$ be the absorption length, $l^*$ be the transport
length at the reflection peak, and $L$ be the sample thickness. When the
thickness is the largest of the three lengthscales, optimal saturation
should correspond to an absorption length that is comparable to the
transport length: $l_\textnormal{abs} \sim l^* \ll L$. If the absorption
length were much smaller than the transport length ($l_\textnormal{abs}
\ll l^* \ll L$), the scattering would be weak and absorption would
dominate. If the absorption length were much larger than the transport
length ($l^* \ll l_\textnormal{abs} \ll L$), multiple scattering would
not be suppressed. If the absorption length is the largest of the three
lengthscales, we obtain the design rule specified above: $l^* \sim L \ll
l_\textnormal{abs}$. Finally, when the transport length is the largest
of the three lengthscales ($l_\textnormal{abs}, L \ll l^*$), we expect
only weak color. The sample will be transparent for $L \ll
l_\textnormal{abs} \ll l^*$ and black for $l_\textnormal{abs} \ll L \ll
l^*$.

%%%%%%%%%%
\section{Conclusions}
%%%%%%%%%%

We have shown that three spectral features determine the structural
color of disordered colloidal materials, and we have established their
origins. The location of the main peak in the reflectance spectrum can
be predicted accurately from a single-scattering model that accounts for
the effective index of the material and its glassy structure, as shown
previously~\cite{magkiriadou}. Our measurements show that near the peak,
most of the light is singly scattered. However, the peak is broader than
predicted by the model because of a peak at a slightly smaller
wavelength that arises from multiple scattering and total internal
reflection. The third spectral feature, an increase in scattering toward
shorter wavelengths, leads to the largest deviation from the model
predictions. We have shown that this multiple scattering is due to the
scattering from individual particles, and its increase is related to the
increase in the single-particle scattering cross-section.

We have also shown that the single-scattering model is a useful tool for
understanding and predicting structural color. The model reproduces the
peak and the shape of the measured single-scattering spectrum. Even
though it does not account for the contribution of multiple scattering
to the reflectance spectra, it can be used to calculate the transport
length, which in turn can be used to predict the onset of multiple
scattering.

To make samples with saturated structural color for applications, it is
necessary to have high scattering at the primary peak and minimal
scattering off peak. The saturation can be maximized by varying the
sample thickness and/or by adding broadband absorbers such as carbon
black to the material~\cite{forster}. In many applications, however, it
may not be possible to add absorbers.  For example, in reflective
displays, absorbing materials lead to heating under illumination.  As
shown here, the single-scattering model provides a way to predict the
optimal thickness, in the absence of absorption, based on the wavelength
dependence of the transport length. Even in the presence of absorption,
an understanding of the physical origins of multiple scattering is
important, since multiple scattering increases the path length of light
and thus the probability of being absorbed~\cite{mccoy_structural_2018}.
Thus, any subsequent models that attempt to predict how a given amount
of broadband absorber affects the color must account for multiple
scattering as well. We leave the development of a model that can predict
the reflection spectrum in the presence of both multiple scattering and
absorption for future work.

\begin{acknowledgments}
  Anna B. Stephenson acknowledges the support of the National Science
  Foundation (NSF) Graduate Research Fellowship Program. This work was
  funded by the Harvard MRSEC under grant number DMR-1420570 and by the
  Xerox University Affairs Committee. In addition, this work was
  performed in part at the Center for Nanoscale Systems (CNS), a member
  of the National Nanotechnology Coordinated Infrastructure Network
  (NNCI), which is supported by the National Science Foundation under
  grant number EECS-1541959. CNS is part of Harvard University. We thank
  Dr.\ Arthur McClelland for helpful discussions on the polarization
  setup and Dr.\ Ming Xiao and Anastasia Ershova for suggestions on our
  manuscript.
\end{acknowledgments}

\appendix
% \setcounter{figure}{0}
% \makeatletter 
% \renewcommand{\thefigure}{A\@arabic\c@figure}
% \makeatother
\begin{figure*}[htbp]
  \centering\includegraphics{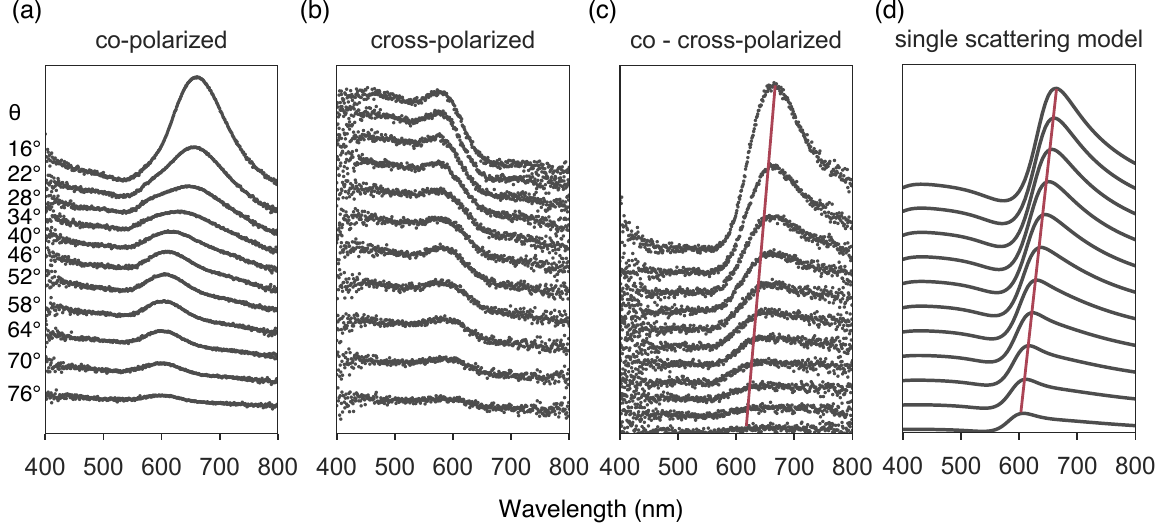}
  \caption{Angle-dependence of polarization spectra. (a) Co-polarized
    reflection spectra. (b) Cross-polarized reflection spectra. (c) Co-
    minus cross-polarized reflection spectra. (d) Reflection spectra
    calculated from the single-scattering model. Each curve in the
    panels corresponds to a different detection angle $\theta$. Spectra
    are offset along the y-axis for clarity. For the measurements in
    (a)-(c), the sample and experimental setup are the same as in
    Fig.~\ref{single_multiple_scat}a. For the calculation in (d), we use
    a volume fraction of 0.53, corresponding to the measured volume
    fraction, a diameter of \SI{280}{nm}, and an effective refractive
    index ranging from 1.322 at \SI{400}{nm} to 1.296 at
    \SI{800}{nm}. \label{angle-dependence}}
\end{figure*}

\section*{Appendix: angle-dependence of polarization spectra}
Measurements of the angle-dependence of the polarization spectra reveal
that the sample is disordered. The primary peak blueshifts as the
detection angle $\theta$ increases (Fig.~\ref{angle-dependence}a, c).
This blueshift matches the peak shift predicted for a disordered
colloidal sample, as calculated from the single-scattering
model.

\bibliography{references}% Produces the bibliography via BibTeX.

\end{document}